\documentclass[aps,prc,amsmath,preprint]{revtex4}

\usepackage{amsmath}
\usepackage{bm}
\usepackage[dvips]{color,graphicx}
\usepackage{epsfig}
\usepackage{dcolumn}

\newcommand{\gapproxeq}{\lower .7ex\hbox{$\;\stackrel{\textstyle >}{\sim}\;$}}
\newcommand{\lapproxeq}{\lower .7ex\hbox{$\;\stackrel{\textstyle <}{\sim}\;$}}

\preprint{JLAB-THY-09-952}

\begin{document}

\title{Duality in Semi-Inclusive Pion Electroproduction}
\author{F. E. Close}
\affiliation{Rudolf Peierls Centre for Theoretical Physics,
	University of Oxford,\\ 1 Keble Road, Oxford, OX1 3NP, UK}

\author{W. Melnitchouk}
\affiliation{Jefferson Lab, 12000 Jefferson~Avenue, 
	Newport News, Virginia 23606, USA}

\begin{abstract}
We explore quark-hadron duality in semi-inclusive pion electroproduction
on proton and neutron targets.  Using the spin-flavor symmetric quark 
model, we compute ratios of $\pi^+$ and $\pi^-$ cross sections for both 
unpolarized and polarized scattering, and discuss realizations of 
duality in several symmetry breaking scenarios.  The model calculations 
allow one to understand some of the key features of recent data on 
semi-inclusive pion production at low energies.
\end{abstract}

\maketitle

\section{Introduction}
\label{sec:Introduction}

Quark-hadron duality in structure functions (also known as Bloom-Gilman
duality \cite{BG}) is a well established empirical phenomenon, relating
measurements in the deep inelastic region to averages over nucleon
resonances (for a review see Ref.~\cite{MEK}).
While a quantitative understanding of its origin in QCD remains elusive,
some insight into the possible realization of duality has recently been
achieved through phenomenological model calculations
\cite{CI01,IJMV,OTHER}.

For inclusive structure functions, Close \& Isgur \cite{CI01} showed
how incorporation of spin-flavor SU(6) symmetry gives quantitative
comparisons between the inelastic structure functions at high energies
for proton or neutron targets and lower energy data covering the
resonance region.
In particular, it was shown that an essential element in the appearance 
of duality was a pattern of constructive and destructive interference
that suppressed multi-quark correlations, leaving only incoherent 
contributions from single quark scattering \cite{CI01}.
In the SU(6) model this was realized by summing over neighboring
positive and negative parity excited states in the $\bm{56}$ and
$\bm{70}$ multiplets, respectively.

When comparing with phenomenology, predictions based on SU(6) symmetry
may be expected to be valid at quark momentum fractions $x \sim 1/3$,
but are known to break down at larger $x$.
Explicit symmetry breaking scenarios, reflecting different patterns in
the flavor-spin dependence of inter-quark forces, were considered in
Ref.~\cite{CM03}, and several of these were found to be consistent
with duality.
The SU(6) breaking analysis extended the range of $x$ up to $x \sim 1$,
and identified implications for the high $Q^2$ behavior of the
$N \to N^*$ transition form factors.

Recently, experimental investigations of duality have been carried out
in charged pion electroproduction from proton and deuteron targets
\cite{HallC,HallCpT}, measuring the ratios of semi-inclusive $\pi^+$
to $\pi^-$ production as a function of $z \equiv E_{\pi}/\nu$,
where $\nu$ is the energy transfer to the target.
The data showed a smooth behavior in $z$, consistent with that
observed by the HERMES \cite{HERMES} and European Muon Collaborations
(EMC) \cite{EMC} at higher energies, raising the question of whether a
similar duality may be at play as that observed in inclusive structure
functions \cite{MEK}.

In fact, a preliminary investigation of duality in semi-inclusive
pion production was made in Ref.~\cite{CI01}, where a factorization
between structure and fragmentation functions was found to hold
by summing over $N^*$ resonances in the quark model.
Ratios of $\pi^+/\pi^-$ yields were calculated for unpolarized cross
sections from protons and neutrons in the SU(6) limit, applicable at
$x \sim 1/3$, and the sum over these was shown to reproduce the
inclusive structure function results.
Duality in semi-exclusive hard pion photoproduction was also considered
in Ref.~\cite{ACW} for fixed center of mass scattering angles.
The applicability of duality for pion photoproduction in specific
kinematic regions was discussed in Ref.~\cite{Hoyer}.
With the advent of the new semi-inclusive data in the resonance region
and beyond \cite{HallC,HallCpT}, it is timely therefore to revisit this
problem by extending the earlier work to the more realistic case of
SU(6) breaking and spin-dependent as well as spin-independent cross
sections.

In this paper we present a detailed account of the semi-inclusive pion
production within a resonance excitation picture, and evaluate the
extent to which quark-hadron duality is realized in the data.
In Sec.~\ref{sec:parton} we review the quark-parton model expectations
for semi-inclusive cross sections ratios in the valence quark region.
The main results of this study, namely the cross sections in terms
of transitions to excited state resonances, are presented in
Sec.~\ref{sec:N*}.
We consider both the spin-flavor symmetric case, as well as several
symmetry-breaking scenarios, and test the validity of duality for
each scenario.
The full set of matrix elements for the transitions
$\gamma N \to N_1^* \to \pi N_2^*$, as well as an explicit example
of a typical matrix element computation, are listed in the Appendices.
Comparisons with recent semi-inclusive pion production data from
Jefferson Lab and elsewhere are discussed in Sec.~\ref{sec:data},
where we generalize our to include non-leading fragmentation.
Finally, some conclusions and ideas for future work outlined in
Sec.~\ref{sec:conc}.

\section{Parton Model Fragmentation}
\label{sec:parton}

If the hadronization process is independent of the target, the 
semi-inclusive cross section can be factorized into a product of a 
parton distribution function describing the hard scattering from a 
parton in the target, and the probability of the struck parton 
fragmenting into a specific hadron, parameterized by the fragmentation 
function.
At the quark level, the semi-inclusive cross section for the production 
of pions from a nucleon target in the valence quark region
($x \gapproxeq 0.3$) is proportional (at leading order in $\alpha_s$) to
\begin{equation}
{\cal N}_N^\pi(x,z)\
=\ e_u^2\ u^N(x)\ D_u^\pi(z)\ +\ e_d^2\ d^N(x)\ D_d^\pi(z)\ ,
\label{eq:Nxz}
\end{equation}
where $e_u\ (e_d) = 2/3\ (-1/3)$ is the $u\ (d)$ quark charge,
$q^N$ is the distribution of quarks $q$ in the nucleon $N$ with 
light-cone momentum fraction $x$, and $D_q^\pi$ is the fragmentation 
function for quark $q$ to produce a pion with energy fraction
$z = E_\pi/\nu$, with $E_\pi$ the pion energy and $\nu$ the energy
transfer to the target.
In both the distribution functions and fragmentation functions we have
suppressed the explicit dependence on the scale $Q^2$.

The fragmentation functions are traditionally separated into favored
and disfavored classes, $D_u^{\pi^+} = D_d^{\pi^-} \equiv D$
and $D_d^{\pi^+} = D_u^{\pi^-} \equiv \overline D$, respectively,
where we have further assumed isospin symmetry in relating the 
fragmentation functions for $\pi^+$ and $\pi^-$.
In the leading fragmentation approximation, valid in the limit $z \to 1$,
the hadronization process is dominated by the favored fragmentation,
and $\overline D \to 0$.
In this approximation the cross sections for proton and neutron targets 
can be written as:
\begin{subequations}
\label{eq:Nqpm}
\begin{eqnarray}
{\cal N}_p^{\pi^+}(x,z) &=& e_u^2\ u(x)\ D(z)\ ,	\\
{\cal N}_p^{\pi^-}(x,z) &=& e_d^2\ d(x)\ D(z)\ ,	\\
{\cal N}_n^{\pi^+}(x,z) &=& e_u^2\ d(x)\ D(z)\ ,	\\
{\cal N}_n^{\pi^-}(x,z) &=& e_d^2\ u(x)\ D(z)\ ,
\end{eqnarray}
\end{subequations}
where the quark distributions are defined to be those in the proton.

Similarly for a polarized target, the spin-dependent cross section can
be written (at leading order) in terms of spin-dependent distribution
functions $\Delta q$ and corresponding fragmentation functions
$\Delta D$,
\begin{equation}
\Delta{\cal N}_N^\pi(x,z)\
=\ e_u^2\ \Delta u^N(x)\ \Delta D_u^\pi(z)\
+\ e_d^2\ \Delta d^N(x)\ \Delta D_d^\pi(z)\ .
\end{equation}
Since the quark spins in a pion must average to zero, for pion 
production one expects the fragmentation functions to be independent
of the polarization of the fragmenting quark, so that
$\Delta D_q^\pi = D_q^\pi$.

If the quark distributions and fragmentation functions factorize as
in Eq.~(\ref{eq:Nxz}), one can construct ratios of $\pi^-$ to $\pi^+$
yields on the neutron and proton which are given by simple ratios of
quark charges,
\begin{equation}
{ {\cal N}_n^{\pi^+} \over {\cal N}_p^{\pi^-} }\
=\ { {\cal N}_p^{\pi^+} \over {\cal N}_n^{\pi^-} }\
=\ {e_u^2 \over e_d^2}\ =\ 4\ .
\end{equation}    
Similarly for the spin-dependent cross sections, one finds the
model-independent relations:
\begin{equation}
{ \Delta{\cal N}_n^{\pi^+} \over \Delta{\cal N}_p^{\pi^-} }\
=\ { \Delta{\cal N}_p^{\pi^+} \over \Delta{\cal N}_n^{\pi^-} }\
=\ 4\ .
\end{equation}    
This will serve as useful consistency checks on the duality between the 
partonic and hadronic descriptions of the semi-inclusive cross sections
in the following section.

\section{Resonance Transitions}
\label{sec:N*}

In this section we describe the semi-inclusive production of pions
using a hadronic basis within the SU(6) quark model, generalizing the
discussion of Refs.~\cite{CI01,CM03} on inclusive structure functions.
In the earlier work \cite{CI01,CM03}, the SU(6) model served as a useful 
framework in allowing one to visualize the underlying principles of 
quark-hadron duality and to provide a reasonably close contact with 
structure function phenomenology.
As pointed out by Close and Isgur \cite{CI01}, duality between structure
functions represented by (incoherent) parton distributions and by a
(coherent) sum of squares of form factors can be achieved by summing
over neighboring odd and even parity states.
In the SU(6) model this is realized by summing over states in the
$\bm{56^+}$ ($L=0$, even parity) and $\bm{70^-}$ ($L=1$, odd parity)
multiplets, with each representation weighted equally.

For semi-inclusive scattering, pion production cross sections are
constructed by summing coherently over excited nucleon resonances
($N_1^*$) in the $s$-channel intermediate state and in the final state
($N_2^*$) of $\gamma N \to N_1^* \to \pi N_2^*$, where both $N_1^*$
and $N_2^*$ belong to the $\bm{56^+}$ and $\bm{70^-}$ multiplets.
Duality is then demonstrated by comparing the hadronic-level results
with those of the parton model in Sec.~\ref{sec:parton}.

To generalized the model to values of $x$ away from $x \sim 1/3$,
where SU(6) symmetry is expected to be valid, we incorporate various
SU(6) breaking mechanisms, along the lines of those discussed in
Ref.~\cite{CM03}.
We recall that at the quark level, explicit SU(6) breaking mechanisms
produce different weightings of components of the initial state wave
function, which in turn induces different $x$ dependences for the spin
and flavor distributions.
On the other hand, at the hadronic level SU(6) breaking in the matrix 
elements leads to suppression of transitions to specific $N_1^*$ and 
$N_2^*$ resonances, while starting from a symmetric SU(6) wave function
for the initial state $N$.

\subsection{SU(6) Symmetric Model}
\label{ssec:SU6}

The amplitudes for the transitions $\gamma N \to N_1^*$ correspond
to the absorption of a transversely polarized photon ($J_z=+1$),
exciting the nucleon to a state $N_1^*$ with total angular momentum
$J_z=1/2\ (3/2)$ described by the helicity amplitude $A_{1/2\ (3/2)}$.
As in Refs.~\cite{CI01,CM03,FEC74}, we assume the interaction operator
to be magnetic spin-flip, $\sum_i\ e_i\ \sigma^+_i$, where $e_i$
is the charge of the $i$-th quark and
$\sigma^+ = (\sigma_x + i \sigma_y)/2$ is the Pauli spin raising
operator, as appropriate in the deep inelastic limit \cite{CG72}.
(This corresponds to the ``$B$'' terms in Ref.~\cite{COT}.)

For the pion emission operator we use leading term from Ref.~\cite{COT}
proportional to $\sum_i\ \tau_i^\pm\ \sigma_{z i}$ for $\pi^\mp$
emission, where $\tau_i^\pm$ is the isospin raising/lowering operator
(for the most general form of the operator see Ref.~\cite{COT}).
The leading operator is in fact general for unpolarized scattering,
but for spin-dependent transitions it implicitly assumes that the
emitted pion is collinear (or with $L_z = 0$), which is strictly valid
only in the $z \to 1$ limit.
For simplicity we retain this form in the present exploratory study,
whose primary aim is to demonstrate the workings of duality in
semi-inclusive scattering.
In addition, the currently available data are for unpolarized cross
sections, and refinements of the model to investigate the polarization
dependence as a function of $z$ can be made in future when new
spin-dependent data become available.

Within this framework, the probabilities of the
$\gamma N \to \pi N_2^*$ transitions can be obtained from
Tables~\ref{tab:2_8_56}--\ref{tab:2_10_70} in
Appendix~\ref{sec:amplitudes} by summing over the intermediate states
$N_1^*$ spanning the $\bm{56^+}$ and $\bm{70^-}$ multiplets.
In Table~\ref{tab:su6} we list the relative spin-averaged
${\cal N}_N^\pi$ and spin-dependent $\Delta{\cal N}_N^\pi$
semi-inclusive cross sections (scaled by a factor $9^2=81$)
for the SU(6) symmetric case for specific $N_2^*$ final states,
together with the sum over the $N_2^*$ states.
In the hadronic language we define the yield as
\begin{eqnarray}
{\cal N}_N^\pi(x,z)
&=& \sum_{N_2^*}
\left| \sum_{N_1^*}
       F_{\gamma N \to N_1^*}(Q^2,M_1^*)\
       {\cal D}_{N_1^* \to N_2^* \pi}(M_1^*,M_2^*)\
\right|^2\ ,
\end{eqnarray}
where $F_{\gamma N \to N^*}$ is the $\gamma N \to N^*$ transition form
factor, which depends on the masses of the virtual photon and excited
nucleon ($M_1^*$), and ${\cal D}_{N_1^* \to N_2^* \pi}$ is a
function representing the decay $N_1^* \to \pi N_2^*$, where $M_2^*$
is the invariant mass of the final state $N_2^*$.

\begin{table}[h]
\begin{tabular}{r|c|c|c|c|c|c}					\hline
\multicolumn{7}{c}{$N_2^*$}					\\
	& $\bm{^2 8,56^+}$
	& $\bm{^4 10,56^+}$
	& $\bm{^2 8,70^-}$
	& $\bm{^4 8,70^-}$
	& $\bm{^2 10,70^-}$
	& sum							\\ \hline
$\gamma p \to \pi^+ N_2^*$\ \
        & 100 (100)  & 32 (--16)  & 64 (64)  & 16 (--8) & 4 (4)
	& 216 (144)						\\
$\gamma p \to \pi^- N_2^*$\ \
	& 0 (0)	     & 24 (--12)  & 0 (0)    & 0 (0)    & 3 (3)
	& 27 (--9)						\\
$\gamma n \to \pi^+ N_2^*$\ \
	& 0 (0)	    & 96 (--48)   & 0 (0)    & 0 (0)    & 12 (12)
	& 108 (--36)						\\
$\gamma n \to \pi^- N_2^*$\ \
	& 25 (25)   & 8 (--4)     & 16 (16)  & 4 (--2)  & 1 (1)
	& 54 (36)						\\ \hline
\end{tabular}
\vspace*{0.5cm}
\caption{Relative spin-averaged (spin-dependent) probabilities
	${\cal N}_N^\pi$\ $(\Delta{\cal N}_N^\pi)$ for the
	$\gamma N \to N_1^* \to \pi N_2^*$ transitions in the SU(6)
	symmetric quark model, after summing over all intermediate
	states $N_1^*$.\\}
\label{tab:su6}
\end{table}

Summing over the $N_2^*$ states in the $\bm{56^+}$ and $\bm{70^-}$
multiplets in Table~\ref{tab:su6}, we arrive at the following ratios
of unpolarized $\pi^-$ to $\pi^+$ semi-inclusive cross sections:
\begin{equation}
{ {\cal N}_p^{\pi^-} \over {\cal N}_p^{\pi^+} }\
=\ {1 \over 8}\ ,\ \ \ \ \ \
{ {\cal N}_n^{\pi^-} \over {\cal N}_n^{\pi^+} }\
=\ {1 \over  2}\ ,
\label{eq:SU6rat}
\end{equation}
while for the corresponding spin-dependent $\pi^-$ to $\pi^+$ ratios
we have:
\begin{equation}
{ \Delta{\cal N}_p^{\pi^-} \over \Delta{\cal N}_p^{\pi^+} }\
=\ -{1 \over 16}\ ,\ \ \ \ \ \ 
{ \Delta{\cal N}_n^{\pi^-} \over \Delta{\cal N}_n^{\pi^+} }\
=\ -1\ .
\end{equation}
For specific $\pi^+$ and $\pi^-$ production, the ratios of spin-dependent
to spin-averaged cross sections are:
\begin{subequations}
\begin{eqnarray}
{ \Delta{\cal N}_p^{\pi^+} \over {\cal N}_p^{\pi^+} }
&=& {2 \over 3}\ ,\ \ \ \ \ \ 
{ \Delta{\cal N}_p^{\pi^-} \over {\cal N}_p^{\pi^-} }\
 =\ -{1 \over 3}\ ,				\\
{ \Delta{\cal N}_n^{\pi^+} \over {\cal N}_n^{\pi^+} }
&=& -{1 \over 3}\ ,\ \ \ \ \ \ 
{ \Delta{\cal N}_n^{\pi^-} \over {\cal N}_n^{\pi^-} }\
 =\ {2 \over 3}\ .
\end{eqnarray}
\end{subequations}
Finally, for ratios of neutron to proton cross sections with either
$\pi^+$ or $\pi^-$ we have:
\begin{subequations}
\begin{eqnarray}
{ {\cal N}_n^{\pi^+} \over {\cal N}_p^{\pi^+} }
&=& { {\cal N}_p^{\pi^-} \over {\cal N}_n^{\pi^-} }\
 =\ {1 \over 2}\ ,		\\
{ {\cal N}_n^{\pi^+} \over {\cal N}_p^{\pi^-} }
&=& { {\cal N}_p^{\pi^+} \over {\cal N}_n^{\pi^-} }\
 =\ 4\ .
\end{eqnarray}
\end{subequations}
This is consistent with the parton model results for ratios of parton
distributions satisfying SU(6) symmetry \cite{CI01,CM03},
$d/u = 1/2$, 
$\Delta u/u = 2/3$,
$\Delta d/d = -1/3$,
$\Delta d/\Delta u = -1/4$,
confirming the validity of duality for the case $\overline D = 0$,
as in Eq.~(\ref{eq:Nqpm}).

Furthermore, the inclusive results of Ref.~\cite{CI01} can be
recovered by summing over the $\pi^+$ and $\pi^-$ channels.
In this case one finds the familiar results
\begin{subequations}
\begin{eqnarray}
{ {\cal N}_n^{\pi^+ + \pi^-} \over {\cal N}_p^{\pi^+ + \pi^-} }
&=& {F_1^n \over F_1^p}\
 =\ {2 \over 3}\ ,		\\
%
%
{ \Delta{\cal N}_p^{\pi^+ + \pi^-} \over {\cal N}_p^{\pi^+ + \pi^-} }
&=& {g_1^p \over F_1^p}\
 =\ {5 \over 9}\ ,		\\
{ \Delta{\cal N}_n^{\pi^+ + \pi^-} \over {\cal N}_n^{\pi^+ + \pi^-} }
&=& {g_1^n \over F_1^n}\
 =\ 0\ ,
\end{eqnarray}
\end{subequations}
again consistent with the parton level expectations.

\subsection{Spin-1/2 Dominance}
\label{ssec:spin}

While the SU(6) symmetric results may be expected to be approximately
valid for structure functions at $x \sim 1/3$, at larger $x$ values
strong deviations from SU(6) symmetry are observed.
These can be correlated with SU(6) breaking effects in the transition
form factors, as well as in hadron masses.
It is well known that spin-dependent forces between quarks, such as from 
one gluon exchange \cite{DGG}, introduce a mass difference between the 
nucleon and $\Delta$.
The same mechanism also leads to an anomalous suppression of the
$N \to \Delta$ transition form factor relative to the nucleon elastic 
at high $Q^2$ \cite{FEC73,Stoler}.

If the characteristic $Q^2$ suppression of the $\Delta$ excitation
is related with the spin dependence, then it may be a feature of all
$S=3/2$ states, namely the $[\bm{^4 10, 56^+}]$ and $[\bm{^4 8, 70^-}]$.
In fact, in the approximation of magnetic coupling dominance, which is 
more accurate with increasing $Q^2$, only $S=3/2$ configurations allow 
non-zero $\sigma_{3/2}$ cross sections.
Suppression of these then automatically gives unity for the polarized
to unpolarized ratios,
\begin{equation}
{ \Delta{\cal N}_N^\pi \over {\cal N}_N^\pi }\
 =\ 1\ .      
\end{equation}

For the unpolarized ratios, including only the $\bm{^2 8}$ and
$\bm{^2 10}$ contributions, the $\pi^-$ to $\pi^+$ ratios for the 
proton and neutron become:
\begin{equation}
{ {\cal N}_p^{\pi^-} \over {\cal N}_p^{\pi^+} }\
=\ {1 \over 56}\ ,\ \ \ \ \ \ 
{ {\cal N}_n^{\pi^-} \over {\cal N}_n^{\pi^+} }\
=\ {7 \over 2}\ .
\label{eq:S1/2rat}
\end{equation}
At the parton level, these results imply the ratio 
$\Delta q/q = 1$ for $q=u$ and $d$, and a flavor-dependent
ratio of down to up quark distributions $d/u = 1/14$.
This is consistent with the findings of the inclusive duality
study in Ref.~\cite{CM03}, confirming the duality between the
parton and hadron level descriptions of the semi-inclusive
scattering process also.

\subsection{Scalar Diquark Dominance}
\label{sec:scalar}

If the mass difference between the nucleon and $\Delta$ is attributed to
spin dependent forces, the energy associated with the symmetric part of
the nucleon wave function will be larger than that of the antisymmetric
component.
At the quark level, this pattern of suppression can be realized with
a spin-dependent hyperfine interaction between quarks,
$\vec S_i \cdot \vec S_j$, which modifies the spin-0 and spin-1
components of the nucleon wave function.
Physically this correlates with a ``diquark'' in a $q(qq)$
representation of the nucleon having a larger mass (energy) when the
spin of the $qq$ pair is 1.
A suppression of the symmetric configuration at large $x$ will then give
rise to a softening of the $d$ quark distribution relative to the $u$,
and leads to the proton and neutron polarization asymmetries becoming
unity as $x \to 1$ \cite{FEC74,FEC73}.

In terms of the SU(6) representations, the $^2${\bf 10} and $^4${\bf 8}
multiplets are in the $\bm{70^-}$, and the $^4${\bf 10} unambiguously
in the $\bm{56^+}$ representation.
However, the spin-1/2 $\bm{^2 8}$ can occur in both the $\bm{56^+}$ and
$\bm{70^-}$ representations, and can be written in terms of symmetric
$\psi_\lambda$ and antisymmetric $\psi_\rho$ components, where
$\psi = \varphi \otimes \chi$ is a product of the flavor ($\varphi$)
and spin ($\chi$) wave functions, and $\lambda$ and $\rho$ denote the
symmetric and antisymmetric combinations, respectively \cite{BOOK}.
In the SU(6) limit one has an equal admixture of both $\rho$ and   
$\lambda$ type contributions, whereas in the present scenario only
the $\rho$ components in the $\bm{56^+}$ and $\bm{70^-}$ multiplets
plays a role.
In particular, since transitions to the (symmetric) spin-3/2 states
($^4${\bf 8}, $^4${\bf 10} and $^2${\bf 10}) can only proceed through
the symmetric $\lambda$ component of the ground state wave function, the
$\rho$ components will only excite the nucleon to $^2${\bf 8} states.
If the $\lambda$ wave function is suppressed, only transitions to
$^2${\bf 8} states will be allowed.

To effect this scenario, the simplest strategy is to sum coherently
the $^2${\bf 8} amplitudes in the intermediate $N_1^*$ states
{\em and} in the final $N_2^*$ states in Tables~\ref{tab:2_8_56} and
\ref{tab:2_8_70}, before squaring the result to get the cross section.
Doing so one finds that $\pi^-$ production from the proton and $\pi^+$ 
production from the neutron are both suppressed:
\begin{equation}
{ {\cal N}_p^{\pi^-} \over {\cal N}_p^{\pi^+} }\
=\ 0\ ,\ \ \ \ \ \ 
{ {\cal N}_n^{\pi^+} \over {\cal N}_n^{\pi^-} }\
=\ 0\ ,
\label{eq:S0rat}
\end{equation}
with the ratio of the unsuppressed yields being:
\begin{equation}
{ {\cal N}_n^{\pi^-} \over {\cal N}_p^{\pi^+} }\
=\ {1 \over 4}\ .
\end{equation}
These results are consistent with parton model results in which
$d/u \to 0$ as $x \to 1$.
Finally, since the $^2${\bf 8} amplitudes have no helicity-3/2
components, the spin-dependent semi-inclusive cross sections will
be identical to the spin-averaged ones, with all polarization
asymmetries unity, again as in the parton level model.

\subsection{Helicity-1/2 Dominance}
\label{ssec:helicity}

The central tenet of quark-hadron duality is the non-trivial
relationship between the behavior of structure functions at large $x$
and that of transition form factors at high $Q^2$. 
At high $Q^2$ perturbative QCD arguments suggest that the interaction of
the photon should be predominantly with quarks having the same helicity
as the nucleon \cite{FJ,GUNION}.
Because the scattering of a photon from a massless quark conserves 
helicity, the helicity-3/2 amplitude $A_{3/2}$ would be expected to be
suppressed relative to the $A_{1/2}$ \cite{BOOK}.

In Ref.~\cite{CM03} duality was demonstrated to exist between parton
distributions at large $x$ and resonance transitions classified
according to quark {\em helicity} rather than spin.
Here we examine whether this duality extends also to the semi-inclusive
case with helicity-1/2 dominance.
Table~\ref{tab:hel} lists the relative weights of the
$\gamma N \to N_1^* \to \pi N_2^*$ transitions (scaled by a factor
$9^2=81$), summed over intermediate states $N_1^*$, in this scenario.
The results are obtained by suppressing all helicity-3/2 amplitudes
$A_{3/2}$ in Tables~\ref{tab:2_8_56}--\ref{tab:2_10_70} for both the
$N_1^*$ and $N_2^*$ states.
The results for the spin-averaged and spin-dependent transitions are
therefore identical.

\begin{table}[h]
\begin{tabular}{r|c|c|c|c|c|c}					\hline
\multicolumn{7}{c}{$N_2^*$}					\\
	& $\bm{^2 8,56^+}$
	& $\bm{^4 10,56^+}$
	& $\bm{^2 8,70^-}$
	& $\bm{^4 8,70^-}$
	& $\bm{^2 10,70^-}$
	&\ sum\							\\ \hline
$\gamma p \to \pi^+ N_2^*$\ \
	& 100	& 8	& 64	& 4	& 4	& 180		\\
$\gamma p \to \pi^- N_2^*$\ \
	& 0	& 6	& 0	& 0	& 3	& 9		\\
$\gamma n \to \pi^+ N_2^*$\ \
	& 0	& 24	& 0	& 0	& 12	& 36		\\
$\gamma n \to \pi^- N_2^*$\ \
	& 25	& 2	& 16	& 1	& 1	& 45		\\ \hline
\end{tabular}
\vspace*{0.5cm}
\caption{Relative strengths of the $\gamma N \to N_1^* \to \pi N_2^*$
	transitions for the $F_1$ or $g_1$ structure function in the
	helicity-1/2 dominance model, after summing over all intermediate
        states $N_1^*$.\\}
\label{tab:hel}
\end{table}

The ratios of $\pi^-$ to $\pi^+$ cross sections for proton and neutron
targets in this scenario are given by:
\begin{equation}
{ {\cal N}_p^{\pi^-} \over {\cal N}_p^{\pi^+} }\
=\ {1 \over 20}\ ,\ \ \ \ \ \ 
{ {\cal N}_n^{\pi^-} \over {\cal N}_n^{\pi^+} }\
=\ {5 \over 4}\ .
\label{eq:hel1/2rat}
\end{equation}
Furthermore, the neutron to proton ratios for $\pi^\pm$ production
are given by:
\begin{equation}
{ {\cal N}_n^{\pi^+} \over {\cal N}_p^{\pi^+} }\
=\ { {\cal N}_p^{\pi^-} \over {\cal N}_n^{\pi^-} }\
=\ {1 \over 5}\ ,
\end{equation}
and for consistency one can also verify that
\begin{equation}
{ {\cal N}_n^{\pi^+} \over {\cal N}_p^{\pi^-} }\
 =\ { {\cal N}_p^{\pi^+} \over {\cal N}_n^{\pi^-} }\
 =\ 4\ .
\end{equation}
These results are consistent with the parton model predictions
for helicity-1/2 dominance \cite{FJ}, which yield $d/u = 1/5$.

Suppression of the helicity-3/2 contributions means that the
spin-dependent semi-inclusive cross sections are identical to
the spin-averaged,
\begin{equation}
{ \Delta{\cal N}_N^\pi \over {\cal N}_N^\pi }\
 =\ 1\ . 
\end{equation}
This again is consistent with the parton model calculations which
predict that $\Delta q/q = 1$ for all quark flavors $q$ \cite{FJ}.

\section{Comparison with Data}
\label{sec:data}

Recently the onset of duality in pion electroproduction was studied
in scattering from proton and deuteron targets \cite{HallC}, with the
missing mass $M_2^*$ of the residual system spanning the resonance
region.
Ratios of $\pi^+$ to $\pi^-$ cross sections were found to exhibit
features reminiscent of parton level factorization of the hard 
scattering and subsequent fragmentation processes, even at relatively 
low energies.
In addition, the ratio of unfavored to favored fragmentation functions
was found to closely resemble that observed at much higher energies
from earlier experiments by HERMES \cite{HERMES} and EMC \cite{EMC}.

As noted in Ref.~\cite{HallC}, some of the qualitative features of
the data could be understood in the simple quark model discussed in
Ref.~\cite{CI01} (and in Sec.~\ref{sec:N*} above).
In particular, for a deuteron target the resonance contributions to
the $\pi^+/\pi^-$ ratio appear in a universal 4:1 ratio when summed
over the $\bm{56^+}$ and $\bm{70^-}$ multiplets, and cancel in the
ratio of unfavored to favored fragmentation functions
$R \equiv \overline D/D$, leaving only a smooth background as expected
at higher energies.
The absence of strong resonant enhancement on top of the smooth
background is indeed one of the notable features of the data 
\cite{HallC}.

Although the lack of prominent $N^*$ structure in the semi-inclusive
spectrum is in accord with duality, a quantitative comparison goes
beyond the simple model discussed thus far.
In our model an operator equivalent to $u \to \pi^+ d$ is sandwiched
between SU(6) states for $N_1^*$ and $N_2^*$, after which the initial
and final states are summed over (see Appendix~\ref{sec:example}).
This mechanism allows only for leading fragmentation to take place,
whereas the data \cite{HallC} clearly show that for $z < 1$ the ratio
$R \neq 0$.
The model can therefore only be applied to the resonance part of the 
semi-inclusive cross section, and is not applicable to the nonresonant
background, which makes up a sizable fraction of the data at $z \ll 1$.

The simplest generalization of this formalism to the case of $R \neq 0$
is to retain the assumption that the pion production operator factors 
from the initial and final $N^*$, but allow the possibility that the
$u$ quark, for instance, can fragment to a $\pi^-$.
An example of such a mechanism would be $u \to M^+ X$, where $M^+$ is
some heavy resonance, followed by the decay $M^+ \to \pi^+\pi^+\pi^-$,
leaving the baryonic state $X$ unaltered.
This can be accommodated phenomenologically by assuming, as in
Ref.~\cite{CI01}, that the ratio $R$ is independent of $N^*$.
The relative contributions from the various $N_2^*$ states to the 
semi-inclusive cross section for the $\overline D$ contribution then
follow immediately by interchanging $\pi^+ \longleftrightarrow \pi^-$
in each column of Tables~\ref{tab:2_8_56}--\ref{tab:2_10_70}.

In the particular case of a deuteron target the ratio of $\pi^+$ to
$\pi^-$ cross sections generalizes to
\begin{equation}
{{\cal N}_d^{\pi^+} \over {\cal N}_d^{\pi^-}}
= {4+R \over 4R+1}\ ,
\label{eq:Npi_d}
\end{equation}
and is in fact the same for each multiplet
$N_2^* = [\bm{^2  8, 56^+}]$,
	$[\bm{^4 10, 56^+}]$,
	$[\bm{^2  8, 70^-}]$,
	$[\bm{^4  8, 70^-}]$
or	$[\bm{^2 10, 70^-}]$,
as well as of course for the sum.
Inverting Eq.~(\ref{eq:Npi_d}), we find the ratio
\begin{equation}
R = { 4 - {\cal N}_d^{\pi^+}/{\cal N}_d^{\pi^-} \over
      4\ {\cal N}_d^{\pi^+}/{\cal N}_d^{\pi^-} - 1 }\ ,
\end{equation}
which clearly vanishes in the limit when
${\cal N}_d^{\pi^+} = 4\ {\cal N}_d^{\pi^-}$.
Interestingly, the 4:1 ratio is realized not only for the SU(6) 
symmetric model, but also for the various SU(6) breaking scenarios 
discussed in Sec.~\ref{sec:N*}.
In order to discriminate between the different symmetry breaking
mechanisms, it is therefore not sufficient to simply extend the 
kinematics to the large-$x$ region with a deuteron target; one must
consider explicitly proton (or neutron) targets, where the different
symmetry breaking models yield very different $\pi^+/\pi^-$
predictions --- see Eqs.~(\ref{eq:SU6rat}), (\ref{eq:S1/2rat}),
(\ref{eq:S0rat}) and (\ref{eq:hel1/2rat}).

Experimentally, the presence of secondary, or unfavored, fragmentation
gives a sizable contribution to the semi-inclusive cross section,
and dilutes the simple predictions for the $\pi^+$ to $\pi^-$ ratios
derived in Sec.~\ref{sec:N*}.
However, even though a quantitative comparison of the ratios may at
present be beyond reach, some general features of the data can
nevertheless be understood within the favored fragmentation scenario.

\begin{figure}
\hspace*{-2cm}\includegraphics[height=15cm]{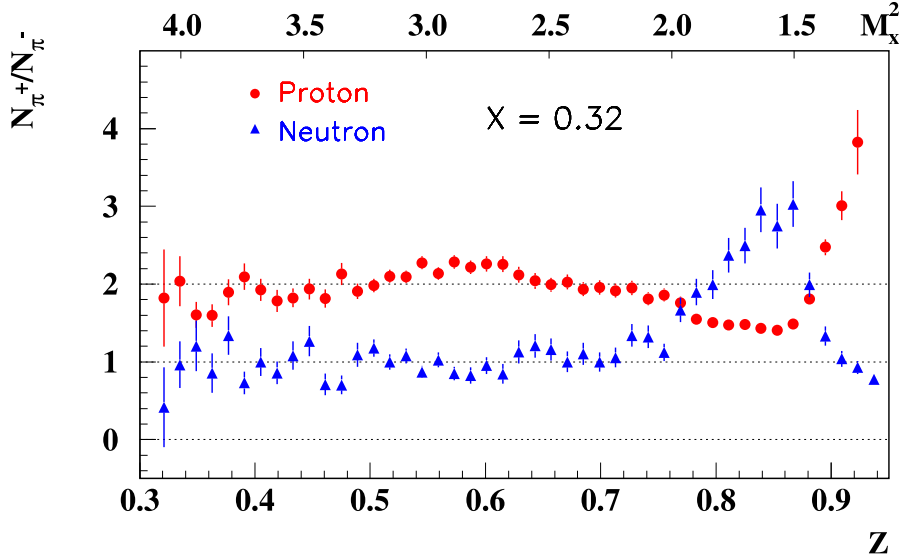}
\vspace*{-8cm}
\caption{(Color online)
	Ratio ${\cal N}^{\pi^+}/{\cal N}^{\pi^-}$ for proton (circles)
	and ``neutron'' (triangles) targets from Jefferson Lab Hall~C
	\cite{HallC,Hamlet}, as a function of $z$
	(and $M_X^2 \equiv M_2^*$, in units of GeV$^2$, upper scale),
	at fixed $x=0.32$ and $Q^2 \approx 2.5$~GeV$^2$.}
\label{fig:data}
\end{figure}

For a proton target, the empirical ratio
${\cal N}_p^{\pi^+}/{\cal N}_p^{\pi^-}$
is found to be $\approx 2$ for $0.3 \lesssim z \lesssim 0.7$,
as seen in Fig.~\ref{fig:data} for fixed $x=0.32$ and
$Q^2 \approx 2.5$~GeV$^2$ \cite{MEK,Hamlet}.
This is to be compared with a $\pi^+/\pi^-$ ratio of 8 predicted
in the SU(6) model when $z \to 1$, Eq.~(\ref{eq:SU6rat}).
A careful examination of the data \cite{Hamlet} at larger $x$ further
reveals a clear trend in which the $\pi^+/\pi^-$ ratio increases with
increasing $x$, consistent with the expectations of the symmetry
breaking scenarios.
For example, the helicity 1/2 dominance model predicts a $\pi^+/\pi^-$
ratio of 20 (Eq.~(\ref{eq:hel1/2rat})), the spin 1/2 dominance model
gives 56 (Eq.~(\ref{eq:S1/2rat})), while the scalar diquark model yields
a divergent ratio (Eq.~(\ref{eq:S0rat})).
Of course to discriminate among these predictions requires data at
larger $x$ ($x \sim 1$) than is currently available, which will be an
important challenge for future experiments.

For deuterium targets, the empirical
${\cal N}_d^{\pi^+}/{\cal N}_d^{\pi^-}$ ratio is found to be around 1.5
\cite{HallC}, compared with the model prediction of 4, in both the SU(6)
symmetric model and in the various symmetry breaking extensions.
As for the proton, the trend of the deuteron data is an increase
of this ratio with increasing $x$.

Since nuclear effects in the deuteron at $x \sim 0.3$ are small
\cite{nuclear}, the $\pi^+/\pi^-$ ratios for a neutron target can
be extracted from the difference of the deuteron and proton data,
as shown in Fig.~\ref{fig:data}.
Here the ${\cal N}_n^{\pi^+}/{\cal N}_n^{\pi^-}$ ratio is close
to unity, and considerably smaller than the proton ratio.
In the SU(6) quark model the expectation for the ratio is 2,
Eq.~(\ref{eq:SU6rat}), so that again the data suggest a dilution
of the primary fragmentation mechanism of $\pi^\pm$ production.

At larger $z$ (or smaller $M_2^*$), the nucleon resonance structures
become more apparent in both the proton and neutron data.
Because secondary fragmentation is suppressed as $z \to 1$, we may
expect the nonresonant background to play a lesser role here and the
predictions of the resonance model of Sec.~\ref{sec:N*} to be more
quantitative.
In the $\Delta$ resonance region, for example, at
$0.8 \lesssim z \lesssim 0.85$ ($M_2^* \approx 1.2-1.3$~GeV),
the proton ratio dips to about 1.5, while the neutron ratio
increases severalfold and becomes larger than the proton.
This behavior is expected from the relative strengths of the
transitions for the $[\bm{^4 10, 56^+}]$ multiplet in
Tables~\ref{tab:su6} and \ref{tab:hel}, in which the neutron
$\to \Delta^0$ transitions are predicted to be an order of magnitude
larger than for the proton $\to \Delta^+$.

For even larger $z$ the proton and neutron ratios are inverted again,
and display rather different behaviors in the limit $z \to 1$,
which corresponds to a nucleon elastic final state, $N_2^* = N$.
Since only $\pi^+$ production is possible from the proton, and $\pi^-$
from the neutron, the ratio of $\pi^+$ to $\pi^-$ rises steeply for
the proton but drops rapidly to zero for the neutron \cite{MEK}.
Again this trend is consistent with the quark model expectations
from Sec.~\ref{sec:N*}, whose predictions should be more robust
in this region.


\section{Conclusion}
\label{sec:conc}

In this study we have shown how the quark-hadron duality that was
successfully demonstrated for inclusive structure functions in the
quark model \cite{CI01} can be extended to semi-inclusive pion
electroproduction, with corresponding patterns of resonances that
are dual to the parton model results.
Having derived the full set of matrix elements for 
$\gamma N \to N_1^* \to \pi N_2^*$, where both $N_1^*$ and $N_2^*$
belong to the $\bm{56^+}$ and $\bm{70^-}$ multiplets of SU(6),
we demonstrated duality by comparing the hadronic-level results,
summed over the $N_{1,2}^*$, with those of the parton model.
In the case of spin-flavor symmetry, these are most immediately
applicable when $x \sim 1/3$ and in the leading fragmentation
approximation, which is valid at large $z$, where secondary
fragmentation is suppressed.

We find that duality is indeed realized even in the broken SU(6) case,
in the sense that sums over resonances are able to reproduce parton
model semi-inclusive cross section ratios.
In Ref.~\cite{CM03} it was shown how duality for inclusive structure
functions could be realized in various SU(6) breaking scenarios;
here we have shown how duality is realized in these scenarios for
semi-inclusive pion electroproduction as $x \to 1$.
The different patterns of resonances $N_2^*$ and ratios of
$\pi^+/\pi^-$ may be used to isolate the dynamics of symmetry
breaking in the $x \to 1$ regime.

Comparisons with data show that this analysis gives a qualitative
description of the features and general trends of the data,
especially at large $z$.
While the results of Ref.~\cite{CI01} were obtained in the leading
fragmentation approximation, which is valid at large $z$,
quantitative comparison with data at intermediate $z$ values
requires including subleading fragmentation, characterized by
non-zero values of the fragmentation function ratio $\overline D/D$.
We have shown how to incorporate the effects of $\overline D/D \neq 0$ 
phenomenologically in a way consistent with duality, although the $z$
dependence is a model-dependent effect that goes beyond the present
work and does not follow from duality alone.

While future precision semi-inclusive studies at large $x$ will
be challenging experimentally, they will be vital in providing
constraints on the flavor and spin dependence of quark interactions
in the nucleon, complementary to those from inclusive measurements.
In this regard a program of polarized semi-inclusive deep-inelastic
scattering to measure spin-dependent cross section ratios would also
be extremely valuable, allowing direct tests of the predictions
presented in this work.

On the theoretical front, quantitative comparison with data at low $Q^2$
and large $x$ and $z$ will require implementation of semi-inclusive
target mass corrections \cite{TMC}, as well as nuclear effects when
considering nuclear data at large $x$ \cite{nuclear}.
For a systematic analysis of the full $z$ dependence of semi-inclusive
cross sections, the present model needs to be extended to include a
microscopic, quark-level description of unfavored fragmentation,
$\overline D/D \neq 0$.
This would then enable predictions to be made for the angular or $p_T$
dependence of semi-inclusive pion production \cite{HallCpT}, which may
reveal further insights into the hadronization process.

\section*{Acknowledgements}

We are indebted to R.~Ent and H.~Mkrtchyan for helpful communications
about the semi-inclusive data from Hall~C at Jefferson Lab \cite{HallC}.
W.M. is supported by the DOE contract No. DE-AC05-06OR23177, under
which Jefferson Science Associates, LLC operates Jefferson Lab.
F.E.C is supported in part by grants from the Science \& Technology
Facilities Council (UK).

\appendix
\section{$\gamma N \to \pi N_2^*$ Amplitudes}
\label{sec:amplitudes}

In this Appendix we present for completeness the full set of 
amplitudes for the transitions $\gamma N \to N_1^* \to \pi N_2^*$,
for $N=p$ and $n$ targets.
In Tables~\ref{tab:2_8_56}--\ref{tab:2_10_70} we give the relative
weights for the helicity amplitudes $A_{1/2}$ and $A_{3/2}$
for individual $N_1^*$ intermediate states and $N_2^*$ final states
spanning the $\bm{56^+}$ and $\bm{70^-}$ multiplets.
We follow the notations of Refs.~\cite{BOOK,FEC74,IK}, but with
a different sign convention for the Clebsch-Gordan coefficients
to that of the Particle Data Group \cite{PDG} (the results for
the cross sections do not depend on the convention however).

\begin{table}[h]        
\begin{tabular}{l|cc|cc|cc|cc}
\multicolumn{9}{c}{$N_2^*=[\bm{^2 8,56^+}]$}			\\ \hline
		& \multicolumn{2}{c|}{$\gamma p \to \pi^+ N_2^*$}
		& \multicolumn{2}{c|}{$\gamma p \to \pi^- N_2^*$}
		& \multicolumn{2}{c|}{$\gamma n \to \pi^+ N_2^*$}
		& \multicolumn{2}{c} {$\gamma n \to \pi^- N_2^*$}\\
\ \ \ $N_1^*$	& $A_{1/2}$		& $A_{3/2}$
		& $A_{1/2}$		& $A_{3/2}$
		& $A_{1/2}$		& $A_{3/2}$
		& $A_{1/2}$		& $A_{3/2}$		\\ \hline
$\bm{^2 8,56^+}$
		& ${5\over 9}$		& 0
		& 0			& 0
		& 0			& 0
		& $-{10\over 27}$	& 0			\\
$\bm{^4 10,56^+}$
		& ${4 \over 27}$	& 0
		& 0			& 0
		& 0			& 0
		& $-{4\over 27}$	& 0			\\
$\bm{^2 8,70^-}$
		& ${4\over 9}$		& 0
		& 0			& 0
		& 0			& 0
		& $-{4\over 27}$	& 0			\\
$\bm{^4 8,70^-}$
		& 0			& 0
		& 0			& 0
		& 0			& 0
		& ${2\over 27}$		& 0			\\
$\bm{^2 10,70^-}$
		& $-{1\over 27}$	& 0
		& 0			& 0
		& 0			& 0
		& ${1\over 27}$		& 0			\\ \hline
\end{tabular}
\vspace*{0.5cm}
\caption{Relative strengths of the helicity 1/2 and 3/2 transition
	amplitudes for $\gamma N \to N_1^* \to \pi N_2^*$ for the
	$N_2^* = [\bm{^2 8,56^+}]$ multiplet.\\}
\label{tab:2_8_56}
\end{table}

\begin{table}[h]        
\begin{tabular}{l|cc|cc|cc|cc}
\multicolumn{9}{c}{$N_2^*=[\bm{^4 10,56^+}]$}			\\ \hline
		& \multicolumn{2}{c|}{$\gamma p \to \pi^+ N_2^*$}
		& \multicolumn{2}{c|}{$\gamma p \to \pi^- N_2^*$}
		& \multicolumn{2}{c|}{$\gamma n \to \pi^+ N_2^*$}
		& \multicolumn{2}{c} {$\gamma n \to \pi^- N_2^*$}\\
\ \ \ $N_1^*$	& $A_{1/2}$			& $A_{3/2}$
		& $A_{1/2}$			& $A_{3/2}$
		& $A_{1/2}$			& $A_{3/2}$
		& $A_{1/2}$			& $A_{3/2}$	\\ \hline
$\bm{^2 8,56^+}$
		& ${2\sqrt{2} \over 9}$		& 0
		& $-{2\over 3}\sqrt{2\over 3}$ 	& 0
		& $-{4\over 9}\sqrt{2\over 3}$	& 0
		& ${4\sqrt{2} \over 27}$	& 0		\\
$\bm{^4 10,56^+}$
		& $-{2\sqrt{2}\over 27}$	& $-{2\over 3}\sqrt{2\over 3}$
		& $-{1\over 9}\sqrt{2\over 3}$	& $-{\sqrt{2} \over 3}$
		& $-{1\over 9}\sqrt{2\over 3}$	& $-{\sqrt{2} \over 3}$
		& $-{2\sqrt{2}\over 27}$	& $-{2\over 3}\sqrt{2\over 3}$\\
$\bm{^2 8,70^-}$
		& $-{2\sqrt{2}\over 9}$		& 0
		& ${2\over 3}\sqrt{2\over 3}$	& 0
		& ${2\over 9}\sqrt{2\over 3}$	& 0
		& $-{2\sqrt{2}\over 27}$	& 0		\\
$\bm{^4 8,70^-}$
		& 0				& 0
		& 0				& 0
		& $-{1\over 9}\sqrt{2\over 3}$	& $-{\sqrt{2} \over 3}$
		& ${\sqrt{2} \over 27}$		& ${1\over 3}\sqrt{2\over 3}$\\
$\bm{^2 10,70^-}$
		& $-{4\sqrt{2}\over 27}$	& 0
		& $-{2\over 9}\sqrt{2\over 3}$	& 0
		& $-{2\over 9}\sqrt{2\over 3}$	& 0
		& $-{4\sqrt{2}\over 27}$	& 0		\\ \hline
\end{tabular}
\vspace*{0.5cm}
\caption{Relative strengths of the helicity 1/2 and 3/2 transition
	amplitudes for $\gamma N \to N_1^* \to \pi N_2^*$ for the
	$N_2^* = [\bm{^4 10,56^+}]$ multiplet.\\}
\label{tab:4_10_56}
\end{table}

\newpage

\begin{table}[h]        
\begin{tabular}{l|cc|cc|cc|cc}
\multicolumn{9}{c}{$N_2^*=[\bm{^2 8,70^-}]$}			\\ \hline
		& \multicolumn{2}{c|}{$\gamma p \to \pi^+ N_2^*$}
		& \multicolumn{2}{c|}{$\gamma p \to \pi^- N_2^*$}
		& \multicolumn{2}{c|}{$\gamma n \to \pi^+ N_2^*$}
		& \multicolumn{2}{c} {$\gamma n \to \pi^- N_2^*$}\\
\ \ \ $N_1^*$	& $A_{1/2}$			& $A_{3/2}$
		& $A_{1/2}$			& $A_{3/2}$
		& $A_{1/2}$			& $A_{3/2}$
		& $A_{1/2}$			& $A_{3/2}$	\\ \hline
$\bm{^2 8,56^+}$
		& ${4 \over 9}$			& 0
		& 0			 	& 0
		& 0			 	& 0
		& $-{8 \over 27}$		& 0		\\
$\bm{^4 10,56^+}$
		& $-{4 \over 27}$		& 0
		& 0			 	& 0
		& 0			 	& 0
		& ${4 \over 27}$		& 0		\\
$\bm{^2 8,70^-}$
		& ${5 \over 9}$			& 0
		& 0			 	& 0
		& 0			 	& 0
		& $-{5 \over 27}$		& 0		\\
$\bm{^4 8,70^-}$
		& 0				& 0
		& 0				& 0
		& 0				& 0
		& $-{2 \over 27}$		& 0		\\
$\bm{^2 10,70^-}$
		& ${1 \over 27}$		& 0
		& 0				& 0
		& 0				& 0
		& $-{1 \over 27}$		& 0		\\ \hline
\end{tabular}
\vspace*{0.5cm}
\caption{Relative strengths of the helicity 1/2 and 3/2 transition
	amplitudes for $\gamma N \to N_1^* \to \pi N_2^*$ for the
	$N_2^* = [\bm{^2 8,70^-}]$ multiplet.\\}
\label{tab:2_8_70}
\end{table}

\begin{table}[h]        
\begin{tabular}{l|cc|cc|cc|cc}
\multicolumn{9}{c}{$N_2^*=[\bm{^4 8,70^-}]$}			\\ \hline
		& \multicolumn{2}{c|}{$\gamma p \to \pi^+ N_2^*$}
		& \multicolumn{2}{c|}{$\gamma p \to \pi^- N_2^*$}
		& \multicolumn{2}{c|}{$\gamma n \to \pi^+ N_2^*$}
		& \multicolumn{2}{c} {$\gamma n \to \pi^- N_2^*$}\\
\ \ \ $N_1^*$	& $A_{1/2}$			& $A_{3/2}$
		& $A_{1/2}$			& $A_{3/2}$
		& $A_{1/2}$			& $A_{3/2}$
		& $A_{1/2}$			& $A_{3/2}$	\\ \hline
$\bm{^2 8,56^+}$
		& ${2 \over 9}$			& 0
                & 0                             & 0
                & 0                             & 0
		& ${4 \over 27}$		& 0		\\
$\bm{^4 10,56^+}$
		& ${2 \over 27}$		& ${2 \over 3\sqrt{3}}$
                & 0                             & 0
                & 0                             & 0
		& $-{2 \over 27}$		& $-{2 \over 3\sqrt{3}}$\\
$\bm{^2 8,70^-}$
		& $-{2 \over 9}$		& 0
                & 0                             & 0
                & 0                             & 0
		& $-{2 \over 27}$		& 0		\\
$\bm{^4 8,70^-}$
		& 0				& 0
		& 0				& 0
                & 0                             & 0
		& ${1 \over 27}$		& ${1\over 3\sqrt{3}}$\\
$\bm{^2 10,70^-}$
		& ${4 \over 27}$		& 0
                & 0                             & 0
                & 0                             & 0
		& $-{4 \over 27}$		& 0		\\ \hline
\end{tabular}
\vspace*{0.5cm}
\caption{Relative strengths of the helicity 1/2 and 3/2 transition
	amplitudes for $\gamma N \to N_1^* \to \pi N_2^*$ for the
	$N_2^* = [\bm{^4 8,70^-}]$ multiplet.\\}
\label{tab:4_8_70}
\end{table}

\newpage

\begin{table}[h]        
\begin{tabular}{l|cc|cc|cc|cc}
\multicolumn{9}{c}{$N_2^*=[\bm{^2 10,70^-}]$}			\\ \hline
		& \multicolumn{2}{c|}{$\gamma p \to \pi^+ N_2^*$}
		& \multicolumn{2}{c|}{$\gamma p \to \pi^- N_2^*$}
		& \multicolumn{2}{c|}{$\gamma n \to \pi^+ N_2^*$}
		& \multicolumn{2}{c} {$\gamma n \to \pi^- N_2^*$}\\
\ \ \ $N_1^*$	& $A_{1/2}$			& $A_{3/2}$
		& $A_{1/2}$			& $A_{3/2}$
		& $A_{1/2}$			& $A_{3/2}$
		& $A_{1/2}$			& $A_{3/2}$	\\ \hline
$\bm{^2 8,56^+}$
		& $-{1 \over 9}$		& 0
		& $-{1 \over 3\sqrt{3}}$ 	& 0
		& ${2 \over 9\sqrt{3}}$		& 0
		& $-{2 \over 27}$		& 0		\\
$\bm{^4 10,56^+}$
		& $-{8 \over 27}$		& 0
		& $-{4 \over 9\sqrt{3}}$	& 0
		& $-{4 \over 9\sqrt{3}}$	& 0
		& $-{8 \over 27}$		& 0		\\
$\bm{^2 8,70^-}$
		& ${1 \over 9}$			& 0
		& ${1 \over 3\sqrt{3}}$		& 0
		& $-{1 \over 9\sqrt{3}}$	& 0
		& ${1 \over 27}$		& 0		\\
$\bm{^4 8,70^-}$
		& 0				& 0
		& 0				& 0
		& $-{4 \over 9\sqrt{3}}$	& 0
		& ${4 \over 27}$		& 0		\\
$\bm{^2 10,70^-}$
		& ${2 \over 27}$		& 0
		& ${1 \over 9\sqrt{3}}$		& 0
		& ${1 \over 9\sqrt{3}}$		& 0
		& ${2 \over 27}$		& 0		\\ \hline
\end{tabular}
\vspace*{0.5cm}
\caption{Relative strengths of the helicity 1/2 and 3/2 transition
	amplitudes for $\gamma N \to N_1^* \to \pi N_2^*$ for the
	$N_2^* = [\bm{^2 10,70^-}]$ multiplet.\\}
\label{tab:2_10_70}
\end{table}

\newpage
\section{Sample Matrix Element Computation}
\label{sec:example}

We provide here an example of a calculation of a typical matrix element
in Tables~\ref{tab:2_8_56}--\ref{tab:2_10_70} in the SU(6) quark model.
To be specific we consider the octet statesm symmetric and antisymmetric
parts are labeled $\lambda$ and $\rho$.
The spin-1/2 octets in the $\bm{56^+}$ and $\bm{70^-}$ multiplets are
written
\begin{eqnarray}
[\bm{{}^2 8, 56^+}]
&=& {1\over\sqrt{2}}
    \left( \phi_\rho\ \chi_\rho\ +\ \phi_\lambda\ \chi_\lambda
    \right)\ ,					\\
{}[\bm{{}^2 8, 70^-}]
&=& {1\over\sqrt{2}}
    \left( \phi_\rho\ \chi_\rho\ -\ \phi_\lambda\ \chi_\lambda
    \right)\ ,
\end{eqnarray}
where $\phi$ and $\chi$ denote the flavor and spin wave functions,
respectively, explicit forms for which can be found in
Refs.~\cite{BOOK,IK}.
Note that in this convention the state with a scalar diquark can be
written as
$\left| \phi_\rho\ \chi_\rho \rangle \right.
= \left\{ \left| \bm{{}^2 8, 56^+} \rangle \right.
	+ \left| \bm{{}^2 8, 70^-} \rangle \right.
  \right\} / \sqrt{2}$.

Consider the process $\gamma N \to N_1^*$, with the nucleon at
rest and the photon along the $\hat z$ axis with $J_z=+1$.
The helicity amplitude $A_{1/2}$ then corresponds to a nucleon
having $J_z=-1/2$, leaving the $N_1^*$ with $J_z=+1/2$.
In the limit of magnetic coupling (corresponding to spin-flip),
the electromagnetic current $J^{\rm em}$ transforms as
$\sum_i\ e_i\ \sigma_i^+$, summed over the three valence quarks.
For a proton target, since the matrix elements of this operator
are unity for the $\phi_\rho\ \chi_\rho$ component and zero for
the $\phi_\lambda\ \chi_\lambda$ components, one has
\begin{eqnarray}
\langle \bm{{}^2 8, 56^+} \left| J^{\rm em}
\right| \bm{{}^2 8, 56^+} \rangle
&=& \langle \bm{{}^2 8, 70^-} \left| J^{\rm em}
    \right| \bm{{}^2 8, 56^+} \rangle\
 =\ 1\ .
\end{eqnarray}

Now consider the decay of a positively charged state $N_1^*$ to a
$\pi^+$ and a neutral baryon $N_2^*$, with both the $N_1^*$ and
$N_2^*$ states in either the $[\bm{^{}2 8, 56^+}]$ or
$[\bm{{}^2 8, 70^-}]$ multiplets.
Assuming that the $\pi^+$ is emitted collinearly along the $\gamma N$
axis, which is valid in the $z \to 1$ limit, the $\pi^+$ emission
operator transforms as $J^\pi = \tau^-\ \sigma_z$.
After summing over the three valence quarks, one has the matrix
elements 
\begin{subequations}
\begin{eqnarray}
\langle \phi_\lambda\ \chi_\lambda \left| \tau^-\ \sigma_z
\right| \phi_\lambda\ \chi_\lambda \rangle &=& {1\over 9}\ ,	\\
\langle \phi_\rho\ \chi_\rho \left| \tau^-\ \sigma_z
\right| \phi_\rho\ \chi_\rho \rangle &=& {1\over 9}\ ,
\end{eqnarray}
\end{subequations}
from which one derives the transition matrix elements
\begin{subequations}
\begin{eqnarray}
\langle \bm{{}^2 8, 56^+} \left| J^\pi \right| \bm{{}^2 8, 56^+}
\rangle
&=& \langle \bm{{}^2 8, 70^-} \left| J^\pi \right| \bm{{}^2 8, 70^-}
    \rangle\
 =\ {5 \over 9}\ ,						\\
\langle \bm{{}^2 8, 56^+} \left| J^\pi \right| \bm{{}^2 8, 70^-}
\rangle
&=& \langle \bm{{}^2 8, 70^-} \left| J^\pi \right| \bm{{}^2 8, 56^+}
    \rangle\
 =\ {4 \over 9}\ .
\end{eqnarray}
\end{subequations}
The product of the matrix elements of $J^{\rm em}$ and $J^\pi$ then
give the entries in Tables~\ref{tab:2_8_56} and \ref{tab:2_8_70} above.
For example, for the transition
$\gamma N \to N_1^* [\bm{{}^2 8, 70^-}]
     \to \pi\ N_2^* [\bm{{}^2 8, 56^+}]$
one has
\begin{eqnarray}
\langle \bm{{}^2 8, 56^+} \left|   J^\pi    \right| \bm{{}^2 8, 70^-}
\rangle
\cdot
\langle \bm{{}^2 8, 70^-} \left| J^{\rm em} \right| \bm{{}^2 8, 56^+}
\rangle
&=& {4 \over 9}\ ,
\end{eqnarray}
as given in Table~\ref{tab:2_8_56}.


\end{document}